
%
%




 at10pt

\def\=>{\Rightarrow}
\def\==>{\Longrightarrow}

 \def\dal{\displaystyle{{\hbox to 0pt{$\sqcup$\hss}}\sqcap}}



\def\lto{\mathop
        {\hbox{${\lower3.8pt\hbox{$<$}}\atop{\raise0.2pt\hbox{$\sim$}}$}}}
\def\gto{\mathop
        {\hbox{${\lower3.8pt\hbox{$>$}}\atop{\raise0.2pt\hbox{$\sim$}}$}}}
%
%
%









\def\hat{\widehat}		

\def\ideq{\equiv}		


\def\interior #1 {  \buildrel\circ\over  #1}     




\def\basisvector#1#2#3{
 \lower6pt\hbox{
  ${\buildrel{\displaystyle #1}\over{\scriptscriptstyle(#2)}}$}^#3}



%
 \let\miguu=\footnote
 \def\footnote#1#2{{$\,$\parindent=9pt\baselineskip=13pt%
 \miguu{#1}{#2\vskip -7truept}}}
%
%

\def\BulletItem #1 {\item{$\bullet$}{#1}}


\def\AbstractBegins
{
 \singlespace                                        
 \bigskip\leftskip=1.5truecm\rightskip=1.5truecm     
 \centerline{\bf Abstract}
 \smallskip
 \noindent	
 } 
\def\AbstractEnds
{
 \bigskip\leftskip=0truecm\rightskip=0truecm
 }

\def\ReferencesBegin
{
 \singlespace					   
 \vskip 0.5truein
 \centerline           {\bf References}
 \par\nobreak
 \medskip
 \noindent
 \parindent=2pt
 \parskip=6pt			
 }

\def\section #1 {\bigskip\noindent{\bf #1 }\par\nobreak\smallskip\noindent}

\def\subsection #1 {\medskip\noindent{\it [ #1 ]}\par\nobreak\smallskip}

\def\eprint #1 {{$\langle$e-print archive: #1$\rangle$}}


%


\message{...assuming 8.5 x 11 inch paper}

\magnification=\magstep1	
\raggedbottom
\hsize=6.4 true in
 \hoffset=0.27 true in		
\vsize=8.7 true in

\voffset=0.28 true in         	

\parskip=9pt
\def\singlespace{\baselineskip=12pt}      
\def\sesquispace{\baselineskip=15pt}      




\phantom{}
\vskip -1 true in
\medskip
\rightline{gr-qc/0003043}
\rightline{SU--GP--00/02--1}
\vskip 0.3 true in

\vfill

\bigskip
\bigskip

\sesquispace
\centerline{ {\bf Indications of causal set cosmology}\footnote{*}%
{To appear in {\it Int. J. Theor. Phys.} as part of the proceedings of
 the Peyresq IV workshop on Quantum and Stochastic Gravity, String
 Cosmology and Inflation held June 28-July 3, 1999 in Peyresq,
 France. }}



\singlespace			        


\bigskip
\centerline {\it Rafael D. Sorkin}
\medskip
\smallskip
 
\centerline {\it Department of Physics, 
                 Syracuse University, 
                 Syracuse, NY 13244-1130, U.S.A.}
\smallskip

\centerline {\it \qquad\qquad internet address: sorkin@physics.syr.edu}

\AbstractBegins 
Within the context of a recently proposed family of stochastic dynamical
laws for causal sets, one can ask whether the universe might have
emerged from the quantum-gravity era with a large enough size and with
sufficient homogeneity to explain its 
present-day
large-scale structure.  In
general, such a scenario would be expected to require the introduction of
very large or very small fundamental parameters into the theory.
However, there are indications that such ``fine tuning'' is not
necessary, and a large homogeneous and isotropic cosmos can emerge
naturally, thanks to the action of a kind of renormalization group
associated with cosmic cycles of expansion and re-contraction.
\AbstractEnds


\sesquispace
\bigskip\medskip

Until as recently as a year ago, it could have been said that we had no
proven method by which to arrive at a dynamical law for causal sets.
That is, the theory remained essentially in a kinematical stage, aside
from some considerations of a very general nature about how a
sum-over-histories might be formulated for causal sets.  What has
changed the situation is the discovery of a family of dynamical laws in
which the ``time-evolution'' of the causal set appears as a process of
stochastic growth [1].  
At a technical level, such a dynamics
may be defined 
in terms of a Markov process 
with a time-varying state-space
 --- a process that might be described as the law of motion
of a ``stochastic spacetime''.  
It turns out that 
relatively little freedom remains, 
once one postulates a dynamics of this kind: 
the picture of sequential growth 
leads almost uniquely 
to the dynamical family of [1] 
provided that one agrees to honor 
the discrete analogs of general covariance and (classical) causality.  
I will not try to summarize these developments 
in any detail here, 
or even to introduce the causal set idea itself.  
For that, the reader is referred to [2] and [1].  
Rather, 
I wish to consider briefly
the possible implications of some of these developments for cosmology.

It is true that the 
``sequential growth dynamics'' found in [1]
are classical (non-quantum), 
and it is true also that 
one does not know at present
whether any of them
leads to something like the Einstein equations, 
or even 
to anything resembling a spacetime at all.  
On the other hand, 
directions in which one might seek 
their quantum generalization
are not hard to discern,
and 
--- still at the classical level ---
there is available at least one plausible guess 
at a choice of growth parameters which might reproduce
something like classical spacetime.  
In these circumstances,
and given also the accumulation of mathematical knowledge concerning at
least one special case of 
these dynamics,
it does not seem out of place
to look for indications of how the theory taking shape might offer its
own solutions to some of the recognized puzzles of cosmology.
Specifically, 
I am thinking of the unexplained ``large numbers''
in cosmology 
related to
the large size of the universe and 
its high degree of homogeneity and isotropy.  
(Lurking behind these issues is the
question of why the cosmological constant $\Lambda$ is so small.  
Causal sets so far have provided at best vague hints of why this should be so,
but they {\it have} led to a prediction [3] 
of fluctuations about $\Lambda=0$, 
and indeed, 
fluctuations of a time-dependent magnitude 
whose predicted value for the current universe is just that
which seems to be indicated by the most recent observations.)

If we suppose that the cosmic microwave radiation we see today is descended
directly from radiation which was present 
at the conclusion of the quantum-gravity era,\footnote{*}
{This assumption is denied in ``inflationary'' scenarios according to
 which all matter visible today was created much later, in a process
 of ``reheating''.}
then we can straightforwardly evolve present conditions back to
describe the universe (as much as we can see of it) 
as it was just after the ``Planck time'', 
by which I mean the time when 
the Hubble parameter $H=\dot{a}/a$ 
was near $1$ in natural units.  
One finds 
(using the $1/a^4$ dependence of the energy density of radiation, 
and barring any conspiracies 
involving a time-varying cosmological constant) 
that 
the temperature at that epoch was also near to unity 
but the radius of curvature was 
some 28 orders of magnitude or more above the Planckian value.  
This ``large number'' 
(which corresponds to the large ratio of the 
present-day
 Hubble radius $1/H$ 
to the 
present-day
 wavelength of the microwave background)
is one for which current theory has no convincing explanation.

Only two ways of obtaining such a large number have seemed appealing:
either derive it from some other large number of the underlying
theory (which then has to be explained in its turn)\footnote{*}
{for example, the ratio of the Planck mass to the Higgs mass}
or relate it to 
some conjunctural (i.e. historical) number of cosmology 
whose large size is not in need of explanation, 
such as the age of the universe 
or the number of cycles of contraction and re-expansion 
it has undergone to date.  
This second way of proceeding is the one to which 
some of the recent causal set results 
lend themselves.

To understand why, one must know that, despite being representable
formally as a Markov process, a sequential growth dynamics exhibits a
long memory, such that the present effective laws of motion are
influenced by past behavior.  (Indeed the process is formally Markovian
only because one includes the entire past in the
stochastically evolving
 ``state''.)  
The passage of time, according to this dynamics consists in
a sequence of ``births'' of new elements of the causal set, each of which comes
into being with a definite set of pre-existing ``ancestor elements''.
The dynamical law is specified by giving the relative
probability of each possible choice of ancestor-set (called ``the
precursor'' in [1]), and this in turn, turns out to be given by
a relatively simple expression depending only on the total size $\varpi$
of the precursor and the size $m$ of its maximal layer,\footnote{**}
{In other language, $\varpi$ is the number of all ancestors and $m$ is
 the number of ``immediate ancestors'' or ``parents''.}
namely
$$
   \lambda(\varpi,m) = \sum\limits_k {\varpi-m \choose k-m} t_k \,,
   \eqno(1)
$$
where $t_0$, $t_1$, $t_2\cdots$ is a sequence of non-negative
``coupling constants'' that completely characterizes the dynamics (and
where $t_0\ideq{1}$).  Notice in this formula how the behavior of the
$n^{th}$ element is influenced not only by the ``contemporaneous coupling
constant'' $t_n$, but by the entire history of $t$'s up to that
``time''.

Now among the possible choices of the $t_n$, 
two may be singled out for special consideration.  
The first choice, 
$$
        t_n = t^n                              \eqno(2)
$$ 
for some fixed $t$ ($0<t<\infty$),
is known as {\it transitive percolation} and describes 
a simplistic, time-reversal invariant dynamics 
in which the future of each element is
independent of its past and of relatively ``spacelike'' regions.
(See [1] and [4] for a more complete definition of
transitive percolation dynamics.)
The second choice, 
$$
      t_n = { t^n \over n! }    \,,            \eqno(3)
$$ 
has been suggested 
as a candidate which might yield spacetimes 
with genuine local degrees of freedom and a more
realistic effective law of motion [1].

Let us consider transitive percolation first, 
since its properties are much better understood.  
One knows in particular that, 
with probability 1, 
the universe it describes undergoes an infinite succession of 
cycles of expansion, stasis and contraction 
punctuated by so called {\it posts} [5], 
each of which serves as the progenitor of 
all the elements born in the next cycle.  
The region issuing from any such post
is independent of what preceded it, 
and has for its effective dynamics
that of {\it originary percolation}, 
which is the same as plain percolation, 
except that no element can be born
without having the post among its ancestors [4].
The size to which the 
region following a post
re-expands 
is governed by the parameter $t$,
or equivalently the probability $p=t/(1+t)$.  
For $t{\ll}1$, 
the universe stops expanding at a ``spatial volume'' of 
not much more than $1/t$, 
whose value therefore would have to exceed (say)
$(10^{28})^3{\sim}10^{84}$
in order to do justice to conditions at the time of the ``big bang'', 
assuming, of course, that
the dynamics of transitive percolation 
is at all relevant to the very early universe.\footnote{*}%
{We will see in a moment why this might be the case.  The number
 $10^{84}$ assumes that a spacelike hypersurface in the continuum
 corresponds to a maximal antichain in the causal set, meaning a maximal
 set of causally unrelated elements.  It assumes also that the spatial
 volume of such a hypersurface is equal, up to a factor of order unity,
 to the cardinality of the corresponding antichain.}
The ``fine tuning'' or ``large number'' problem is then 
why $t$ should have such a small magnitude, 
rather than a value near unity.  

It is here that the memory effects embodied in (1) enter.  
Let us suppose for definiteness that 
the true dynamics is given by $t_n=t^n/n!$,
and let us also suppose,
for the sake of argument, 
that an infinite number of posts will occur for this dynamics as well.   
What then will be the effective dynamics
for the portion of the causal set 
following some given post?
(I'll call this portion the ``current era''.)
Let $e_0$ be the post and let it have $N_0$ elements to its past 
($N_0$ ancestors).  
Then, by definition, an element $x$ born in the current era with
$\varpi$ {\it current} ancestors (including $e_0$) will have in reality
$\varpi+N_0$ ancestors in the full causal set.  On the other hand, its
number of parents (maximal elements of past($x$)) will be unaffected by
the region 
preceding
$e_0$, since the presence of $e_0$ prevents
any element in that region from being an immediate ancestor of $x$.  For
the region, future($e_0$), we thus acquire an {\it effective dynamics}
described by weights $\hat\lambda(\varpi,m)$ 
related to the fundamental weights $\lambda(\varpi,m)$    
by the simple equation  
$$
    \hat\lambda(\varpi,m) = \lambda(\varpi+N_0,m)  \,. \eqno(4)
$$

Each cosmic cycle thus acts to renormalize the coupling constants for
the next cycle, and the dynamics in any given cycle differs from the
original or ``bare'' dynamics by the action of this cosmological
``renormalization group''.  
It turns out that, 
when expressed as a
transformation of the elementary coupling constants $t_n$,
this action is very simple.
For $N_0=1$ we have
$$
   \hat{t}_n = t_n + t_{n+1}    \eqno(5)
$$
and for $N_0=2,3,4,\cdots$ we just iterate this transformation $N_0$ times.  
(For defining the
dynamics, only the ratios of the $t_n$ matter.  
Hence, 
the $t_n$ lie in a projective space, and
(5), though it appears linear, is really a projective mapping).
Equation (5) seems so simple that 
one could hope to analyze it fully,
finding in particular all the attractors 
and their ``basins of attraction''.  
Potentially such an analysis could pick out as favored
dynamical laws those to which the universe tends to evolve under the
action of the ``cosmic renormalization group''.  
For now, 
we can note [6]
that the only fixed
points of (5) are those of the percolation family,  $t_n=t^n$.     
(proof:  In order that ratios $t_n:t_m$ not be altered by (5), it is
necessary and sufficient that $\hat{t}_n=ct_n$ for some constant $c$.  
But this holds iff $t_{n+1}=t_n t$ with $t=c-1$.)

In [6], Djamel Dou has studied the action 
of this cosmic renormalization group 
on (3), 
as well as on some other choices of the $t_n$ which can be
regarded as simple ``deformations'' of (2),
like $t_n={t^n}\,{p!}\,{n!}/{(n+p)!}\,$.
For the latter cases
he finds that the ``renormalization group flow'' defined by (5) 
leads back to the fixed point set (2),
indicating that percolation is to some degree an ``attractor''
in the space of all dynamics.
For the former case, 
the story is more interesting.  
In the limit of large $N_0$, 
and for $m^2{\ll}N_0t$, $\varpi{\ll}N_0$, 
one finds that $\hat\lambda(\varpi,m)$ 
corresponds to percolation (2) 
{\it with an $N_0$-dependent parameter $t$} 
given by 
$$
        \hat{t} = \sqrt{t/N_0}   \eqno(6)
$$
The effective dynamics is thus once again transitive percolation, 
but only for a limited time,\footnote{$^\dagger$}
{The initial phase of effective percolation could not last forever.  If
 it did, we could prove that another post would occur, whereafter, by
 (6), we'd have to have percolation with a {\it smaller} $t$,
 contradicting our original assumption.}
and with an effective parameter $t$ that diminishes from one cosmic
cycle to the next.

Now, the germ of a resolution to our cosmological puzzles is contained in
these results.  
Let us adopt the cosmology of (3) 
with its single free parameter 
taken to be a number of order unity
(i.e. no ``fine-tuning''), 
and let us assume that repeated posts occur.  
After each post, 
the ensuing cosmological cycle 
will begin with a stage governed by 
the dynamics (2) 
with a parameter $t=\hat{t}$ 
which diminishes rapidly
from cycle to cycle.  
During each such stage, 
the causal set will expand 
to a spatial volume of at least $O(\hat{t}^{-1})$,
a magnitude which increases rapidly from cycle to cycle.  
Moreover, 
it is not difficult to see that
the earliest portion of this percolation stage 
(that for which $\hat{n}\ll\hat{t}^{-1}$) 
will be a phase of exponential tree-like growth
(a tree being a poset in which every element but the first has precisely
one parent.)\footnote{*}
{Computer simulations confirm this tree-like character, and also confirm
 the deduction [7] that its ``average branching number'' is near
 to two (i.e. the number of children per element, averaged over all the
 elements is about two at any fixed stage of the growth process).} 
At the conclusion of each tree-like phase, we will have a
homogeneous\footnote{**}
{and also isotropic, to the extent that the causal set is sufficiently
 like a manifold that this term has meaning.}
universe with a ``spatial volume'' 
that grows larger with each successive cycle.  
In other words, 
by waiting long enough,
we will automatically obtain conditions very like 
those needed for the ``big bang'' in whose aftermath we live.
The ``unnaturally'' large size 
with which spacetime began in our particular phase of expansion 
would then reflect nothing more than the fact that 
a sufficiently great number of causal set elements 
had accumulated in previous cosmic cycles.  


\bigskip\noindent
Before concluding, 
I would like to thank Chris Stephens and Alan Daughton
for numerous early conversations about the cosmology of percolation
dynamics.
This research was partly supported by NSF grant PHY-9600620 
and by 
a grant from the Office of Research and Computing of Syracuse University.

\ReferencesBegin

[1] 
David P.~Rideout and Rafael D.~Sorkin, ``A Classical Sequential Growth 
 Dynamics for Causal Sets'',
 {\it Phys. Rev.} {\bf D61}:024002 (2000),
 \eprint{gr-qc/9904062} .

[2]		
L.~Bombelli, J.~Lee, D.~Meyer and R.D.~Sorkin, ``Spacetime as a causal set'', 
  {\it Phys. Rev. Lett.} {\bf 59}:521-524 (1987);

R.D.~Sorkin, ``Spacetime and Causal Sets'', 
   in J.C. D'Olivo, E. Nahmad-Achar, M. Rosenbaum, M.P. Ryan, 
       L.F. Urrutia and F. Zertuche (eds.), 
   {\it Relativity and Gravitation:  Classical and Quantum,} 
   (Proceedings of the {\it SILARG VII Conference}, 
    held Cocoyoc, Mexico, December, 1990), pages 150-173,
   (World Scientific, Singapore, 1991);

David D.~Reid, ``Introduction to causal sets: an alternate view of
  spacetime structure''
  \eprint{gr-qc/9909075} .

[3]
 R.D.~Sorkin, 
``First Steps with Causal Sets'', 
  in R. Cianci, R. de Ritis, M. Francaviglia, G. Marmo, C. Rubano, 
     P. Scudellaro (eds.), 
  {\it General Relativity and Gravitational Physics,} 
   (Proceedings of the Ninth Italian Conference of the same name, 
     held Capri, Italy, September, 1990), pp. 68-90
  (World Scientific, Singapore, 1991);

R.D.~Sorkin, 
``Spacetime and Causal Sets'', 
     in J.C. D'Olivo, E. Nahmad-Achar, M. Rosenbaum, M.P. Ryan, 
              L.F. Urrutia and F. Zertuche (eds.), 
    {\it Relativity and Gravitation:  Classical and Quantum} 
    (Proceedings of the {\it SILARG VII Conference}, 
      held Cocoyoc, Mexico, December, 1990), 
    pages 150-173
    (World Scientific, Singapore, 1991);

R.D.~Sorkin,
``Forks in the Road, on the Way to Quantum Gravity'', talk 
   given at the conference entitled ``Directions in General Relativity'',
   held at College Park, Maryland, May, 1993,
   {\it Int. J. Th. Phys.} {\bf 36}: 2759--2781 (1997)   
   \eprint{gr-qc/9706002} .



[4]
David P.~Rideout and Rafael D.~Sorkin,
``Evidence for a continuum limit in causal set dynamics''
 (in preparation); see also reference [1].
%

[5] 
Noga Alon, B{\'e}la Bollob{\'a}s, Graham Brightwell, and Svante Janson,
``Linear extensions of a random partial order'',
{\it Ann. Applied Prob.} {\bf 4:} 108-123 (1994).

[6] 
Djamel Dou,			
 ``Causal Sets, a Possible Interpretation for the Black Hole
 Entropy, and Related Topics'', 
 Ph.~D. thesis (SISSA, Trieste, 1999).

[7] 
Alan Daughton, Rafael D.~Sorkin and C.R.~Stephens,
``Percolation and Causal Sets: A Toy Model of Quantum Gravity''
(in preparation).

\end


(prog1    'now-outlining
  (Outline 
      "
     "......"
      "
   "\\\\message" 
   "\\\\section"
   "\\\\appendi"
   "\\\\Referen"	
   "\\\\Abstrac" 	
      "
   "\\\\subsectio"
   ))